\documentclass[conference]{IEEEtran}
\usepackage[usenames,dvipsnames]{xcolor}
 
\usepackage{algorithm}
\usepackage{algorithmic}
\usepackage{amssymb}
\usepackage{xspace}
\usepackage{comment}
\usepackage{cite}
\usepackage{subfigure}
\usepackage{todonotes}
\usepackage{amsmath,tabularx}
\usepackage{url}

\usepackage{graphicx}

\usepackage{listings}
\newcommand{\name}{DKVF\xspace}

\newcommand{\protobuf}{protobuf\xspace}
\newcommand{\cm}{Cluster Manager\xspace}
\def\code#1{\texttt{#1}}

\begin{document}
\title{DKVF: A Framework for Rapid Prototyping and Evaluating Distributed Key-value Stores}


\author{
\IEEEauthorblockN{Mohammad Roohitavaf and Sandeep Kulkarni}
\IEEEauthorblockA{Computer Science and Engineering Department\\
Michigan State University,
East Lansing, MI, USA\\
Email: \{roohitav, sandeep\}@cse.msu.edu}
}
\maketitle

\begin{abstract}
We present our framework DKVF that enables one to quickly prototype and evaluate new protocols for key-value stores and compare them with existing protocols based on selected benchmarks. 
Due to limitations of CAP theorem, new protocols must be developed that achieve the desired trade-off between consistency and availability for the given application at hand. Hence, both academic and industrial communities focus on developing new protocols that identify a different (and hopefully better in one or more aspect) point on this trade-off curve.
While these protocols are often based on a simple intuition, evaluating them to ensure that they indeed provide increased availability, consistency, or performance is a tedious task. Our framework, DKVF, enables one to quickly prototype a new protocol as well as identify how it performs compared to existing protocols for pre-specified benchmarks. 
Our framework relies on YCSB (Yahoo! Cloud Servicing Benchmark) for benchmarking. 
We demonstrate DKVF by implementing four existing protocols --eventual consistency, COPS, GentleRain and CausalSpartan-- with it. We compare the performance of these protocols against different loading conditions. We find that the performance is similar to our implementation of these protocols from scratch. And, the comparison of these protocols is consistent with what has been reported in the literature. Moreover, implementation of these protocols was much more natural as we only needed to translate the pseudocode into Java (and add the necessary error handling). Hence, it was possible to achieve this in just 1-2 days per protocol. 
Finally, our framework is extensible. It is possible to replace individual components in the framework (e.g., the storage component).

\end{abstract}

\section{Introduction}
Key-values stores, together with other forms of NoSQL storage systems, have gained popularity in recent years due to their advantages over relational databases for modern workloads. The schemaless approach of NoSQL databases has made them a good choice for today's web applications with changing requirements. The need for greater scalability for very large datasets and very high write throughput has led to the ever-increasing use of NoSQL databases for big data and real-time web applications \cite{book}.  

With the huge amount of data and very high query throughput produced by a large number of users across the world, storing data in a single machine does not work for any major business. Thus, we have to distribute the data across several machines. When we distribute our data, an important challenge is the consistency between different copies (i.e., replicas) of the data. There is an inherent trade-off between consistency and availability/performance \cite{cap}. Different levels of consistency come with different levels of availability/performance overhead. Even to achieve a certain level of consistency, two different protocols may have different levels of overhead. 

In general, this suggests that developers need to develop new protocols to improve performance, provide higher levels of consistency, reduce communication requirements, reduce storage requirements, and so on. When the developers intuitively identify a new approach to design such a protocol, the natural question that arises is how to evaluate the new protocol by comparing it with different existing protocols. Generally, the concept of the new protocol is often a simple innovation/intuition (e.g., explicitly keep track of dependencies as in \cite{cops}, using time as a way to decide when keys should be made visible as in \cite{gentleRain}) but its evaluation is more complicated. Distributed data stores are complex systems which makes an accurate analytical performance evaluation infeasible for them. A more practical option is experimental performance evaluation via benchmarking a prototype running the protocol. 

One way to have a prototype is building it from scratch. This approach has the advantage of maximum flexibility. However, building our prototype from scratch may take a long time that can slow down the research or development. Furthermore, if the protocol suffers from some undesirable properties (e.g., low performance), a substantial amount of development time is wasted. Another important obstacle is that this approach especially makes the comparison to other protocols hard. Imagine that we want to compare our new protocol to several other existing systems, developed by other groups. Since each of them are implemented in a different code base, we need to implement other protocols with the same code base as ours to have a fair comparison which requires more time.

Another approach is to create our prototype by modifying an already existing system. There are many open-source NoSQL databases that can be modified for prototyping purposes. An important advantage of this approach is that by building a system on top of a tested system, we can benefit from all of its good features, and save time. However, modifying an existing system has its own disadvantages. The most important problem is the lack of flexibility. Although we can always change the code of an open-source data store, to change a system correctly, we need to understand a possibly massive implementation thoroughly, that may take even more time than creating a prototype from scratch. In addition, by changing an existing product, we may lose the advantage of reusing some of its components which was the whole purpose of using an existing product. For instance, suppose an existing system uses a certain replication policy. If the replication policy of our protocol is different, we have to change the whole replication mechanism of the underlying system.

Another problem of changing an existing product is the problem of being locked by that product. For instance, suppose that we have implemented a prototype to evaluate an algorithm for causal consistency by forking from a current system like Cassandra \cite{cassandra}. If in future we are interested to see how our algorithm would perform if we used another system, say Voldemort \cite{voldemort}, we have no choice but building another system based on Voldemort as well. That would be especially necessary if we want to compare our algorithm with another system based on Voldemort. 

In addition to the implementation of a protocol, running experiments is also another burden. Different research groups may evaluate their systems in different ways making comparisons unfair. Yahoo! Cloud Serving Benchmark (YCSB) \cite{ycsb} is a good candidate for a unified way of comparing different storage systems. The YCSB drivers required for benchmarking with YCSB are already available for many systems. Although YCSB helps us to benchmark our system, writing the driver, running clusters and clients on several machines, obtaining, and aggregating the results is a task that we have to do everything we want to evaluate a new protocol. 

In this paper, we introduce Distributed Key-Value Framework (\name) that allows protocol designers to quickly create prototypes running their protocols to see how they work in practice. We want to help researchers to only focus on their high-level protocol and let the \name do all the lower-level tasks. For instance, consider the GentleRain protocol proposed in \cite{gentleRain}. The server side of this protocol is only 31 lines of pseudocode provided in Algorithm 2 of \cite{gentleRain}. However, to have a prototype running this protocol, we need to write hundreds of lines of code to handle lower-level tasks that are independent of the protocol. Our goal is to provide a framework that helps researchers to create their prototypes by writing codes that are very close to the pseudocodes that they publish in their research papers. We believe this framework together with a toolset that helps us to run experiments can significantly save time in implementing and benchmarking new protocols. We hope our framework expedites the research on the field. 

Followings are the advantages of our framework: 

\begin{itemize}
\item The framework allows us to easily define our protocol in a high-level abstraction with an event-driven approach. Specifically, we can define our protocol as a set of event handlers which is the same way as researchers typically present their protocols in their papers. It makes the code much more clear, and reduces the number of lines of code that protocol designers need to write.

\item The clear separation of concerns that the framework provides expedites debugging the system, and improves maintainability of our code. 

\item We can easily compare any two protocols that are implemented on top of the framework, as both of them are implemented with the same code base. 

\item We provide the implementation of four protocols with this paper. Adding other protocols to the repository is part of the future work. Also, other groups can add their protocols to the repository making them publicly available. Having a library of protocol implementations allows researchers to easily compare their protocols to previous ones. 

\item We can easily change the storage engine of our key-value store without changing the logic of our higher consistency protocol. This makes comparison easy, as we can use the storage engine that another system is built on for comparison purposes. 

\item The framework and its toolset streamline the use of YCSB for benchmarking protocols. It encourages researchers to use a standardized framework for benchmarking instead of performing experiments in individual ways.

\item The framework comes with a command line application called Cluster Manager that lets us conveniently run a cluster over the network. Using Cluster Manager, we can easily run a cluster on cloud systems such as Amazon Web Services (AWSs) on Windows or Linux instances. It allows us to monitor connections, network latencies, current load on nodes, and so on. 

\item Cluster Manager also lets us specify a set of experiments to benchmark the system. it takes care of running YCSB clients, collecting, and aggregating the results.

\item  The framework is also accompanied by a graphical tool called Cluster Designer that lets us easily define our cluster and experiments. We can visually create a graph of servers and clients, and define different workloads to run on the cluster. 

\end{itemize}
The rest of the paper is organized as follows: In Section \ref{sec:back}, we provide a quick background on distributed key-value stores and the problem of consistency. In Section \ref{sec:overview}, we review the overall structure of \name and components that protocol designers need to write. Next, in Section \ref{sec:usage}, we explain how to implement a prototype with \name in details. In Section \ref{sec:ycsb}, we focus on using YCSB to benchmark prototypes created by \name. We introduce \name tools in Section \ref{sec:tools}. We provide our experimental results and analysis in Section \ref{sec:results}. Finally, we provide our conclusion and future work in Section \ref{sec:con}.
\section{Background}
\label{sec:back}
In this section, we briefly provide an overview of distributed key-value stores. We also very briefly explain consistency protocols that we have implemented with \name in this paper. 
\subsection{Distributed Key-value Stores}
Key-value stores provide a simple abstraction to store and retrieve our data. Each key-value store is a set of $\langle key, value \rangle$ pairs. A key-value store has two basic operations: $PUT(k, v)$ and $GET(k)$. $PUT(k, v)$ writes a new version with value $v$ for data item with key $k$, and $GET(k)$ reads the value of the data item with key $k$. A key-value store can store multiple versions for each key, or store only one version for each key. In the single-version type, any time that we update the value, we overwrite the previous version, while in the multi-version type we keep a version chain for each key. Key-value stores are the basis of document-oriented databases. These databases can handle one-to-many relations more efficiently than relational databases. Specifically, we can encode all pieces of data related to a key as a JSON (or XML) document, and store it as the value for that key in our key-value store. This approach provides better \textit{locality} than multi-table schemas that are normally used in relational databases \cite{book}. 

The schemaless data model of document-oriented databases gives us the freedom from strict schema at the write time, and lets the application interpret the structure of the encoded value at the read time. This also eliminates the need for awkward object-relational mapping layers that we usually need for relational databases. Specifically, we can easily encode/decode fields of an object to/from documents that we store in our database. Many new storage systems such as MongoDB \cite{mongoDB}, RethinkDB \cite{rethinkDB}, CouchDB \cite{couchDB}, and Espresso \cite{espresso} support document-oriented data model. 

Typically, for practical systems, to increase the performance, we must distribute our key-value store. Two main techniques to distribute the data are partitioning and replication. Partitioning (also known as sharding) allows us to store our data in more than one machine. Specifically, we can divide the key space into several parts, and store each part on a different machine. Usually, we partition the data in a way that each key is assigned to exactly one partition. Partitioning increases the scalability of our system, as we do not need to fit the whole key space on a single machine. It also enables us to scale our query throughput by adding more nodes \cite{book}. 

Replication, on the other hand, improves durability, availability, and performance. Specifically, by keeping multiple copies of data in several replicas, we can increase the durability of our data. It also improves the availability of the system, as if a replica fails, other replicas can serve the clients. Moreover, using replication, we can keep data geographically close to the clients, thereby reducing the network latency. Because of these benefits, geo-replicated data stores have become an important building block of today's Internet services \cite{cops}.

\subsection{Consistency}
\label{sec:back:consistency}
When we copy the same data across several replicas, we need to make sure that all clients see a consistent view of the data. Different levels of consistency are defined for distributed data stores. Through the light of the CAP theorem \cite{cap}, we know there is an inherent trade-off between availability and consistency. The trade-off also exists for performance and consistency \cite{pacelc}. Specifically, the stronger is a consistency model, the higher is its performance overhead. Even for a certain consistency model, it matters how we achieve it, i.e., two different protocols that achieve equal levels of consistency do not necessarily have the same performance. Thus, anytime that we come up with a new consistency protocol, we need to evaluate its overhead. The goal of \name is to help protocol designers to create a prototype key-value store running their protocols for evaluation purposes. To show the effectiveness of \name, we have implemented four protocols using \name that we briefly review here. 

The first protocol is the eventual consistency protocol. Eventual consistency requires that two connected replicas must finally converge to the same state in the absence of new writes \cite{ec}. We can achieve eventual consistency as follows: Each replica sends new updates that occur in it to other replicas. The receiving replicas simply apply the new updates as they receive them. To converge to the same values, however, all replicas must follow the same rule in applying the updates. Specifically, they must order the versions of a key in the same way. Dynamo \cite{dynamo} and Cassandra \cite{cassandra} are two examples of data stores that provide eventual consistency. 

In addition to the eventual consistency, we also consider three causal consistency protocols. These protocols guarantee that when a version is visible to a client, all of its causal dependencies\cite{causalSpartan} are also visible. COPS \cite{cops} is one of the first protocols for causally consist distributed key-value stores. It guarantees causal consistency by explicitly tracking causal dependencies of a version. Specifically, we keep track of versions that a client reads. Then, once the client writes a value for some key, we consider all the versions that the client has read as the causal dependencies of the new version being written by the client. Each replica sends any new update done by itself to other replicas. When a replica receives a replicate message, it does not make it visible to its clients until it made sure that all of the dependencies of the version are visible in the replica. Since, inside each replica we have multiple partitions, we have to send \textit{dependency check} messages to other partitions to check the dependency. This explicit dependency tracking can affect the performance of the system. 

GentleRain \cite{gentleRain} is another protocol that we implement using \name. To avoid explicit dependency checking mechanism of COPS, GentleRain uses an \textit{implicit} dependency tracking via synchronized physical clocks. It assigns each version a timestamp which is the value of the physical clock at the time of the write. GentleRain assigns timestamp in such a way that if version $v_1$ depends on version $v_2$, the timestamp assigned to $v_1$ is greater than $v_2$. To satisfy this requirement, GentleRain may need to delay some PUT operations, if the physical clocks of the partitions are not perfectly synchronized. Next, each replica calculates a Global Stable Time (GST) which is a value such that any version with timestamp smaller than it is visible inside the replica. Partitions inside a replica need to communicate with each other to calculate GST. Now, when a client asks for a key, we give the client  the most recent version of the key that has a timestamp smaller than the GST. This guarantees that versions are visible only after their causal dependencies.

The delay in PUT operations that GentleRain requires can affect the write throughput of the key-value store. This is especially important for modern workloads that require very high write throughput \cite{book}. This issue is made worse in presence of query amplification where a single end user request translates to many internal operations. In this situation, any delay in any of internal queries increases the final response time perceived by the client, and affects the end user experience \cite{facebook}\cite{causalSpartan}. CausalSpartan solves this issue by replacing physical clock with Hybrid Logical Clock (HLC) \cite{hlc}. HLC as the name suggests, has a hybrid nature that includes the benefits of both logical clock \cite{lamport} and physical clock. At one hand, it provides logical clock property (i.e., if $f$ depends on $e$, then HLC timestamp of $f$ is larger than $e$). On the other, unlike logical clock that has no relation to the physical clock, HLC values are very close to the physical clock. Also, like physical clock, HLC advances spontaneously. Using HLCs, we do not need to force any delay for PUT operations, thereby decreasing the response time and improving the throughput. CausalSpartan also improves update visibility latency that allows us to make remote update visible sooner than GentleRain in case of having a slow replica in the system. To lower update visibility latency, however, CausalSpartan increases the size of metadata (proportional to the number of replicas).
\section {Overview of DKVF}

\label{sec:overview}

\name is written in Java. Each key-value store created based on DKVF has two sides: 1) a server side, and 2) a client side. The server side (respectively, client side) extends server side (respectively, client side) of DKVF by implementing the respective abstract methods and adding new methods required for the protocol at hand.

\begin{figure}
\begin{center}
\includegraphics[scale=0.32]{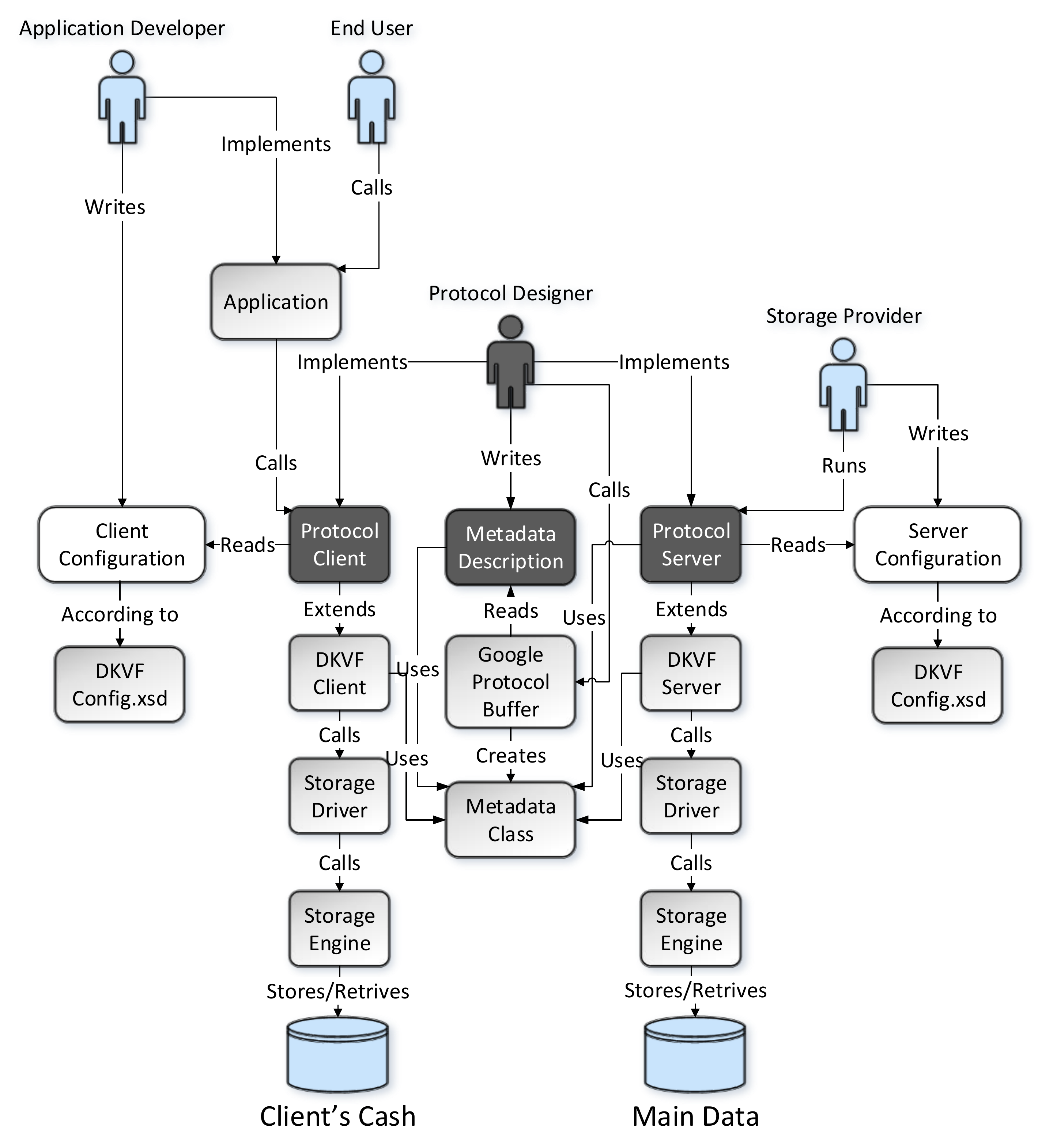}
\caption{Typical usage of DKVF}
\label{fig:overall}
\end{center}
\end{figure}

When we create a new protocol, in addition to actual data consisting  of the key-value pairs, we will likely need to store some metadata with each record. For example, we may need to store a timestamp with each version, or we may need to store the ID of the replica where the version has been written. Each protocol requires its own metadata.  \name relies on Google Protocol Buffers \cite{protobuf} (referred to as \protobuf from now on) for marshalling/unmarshalling data for storage and transmission. An important advantage of \protobuf is its convenience for the protocol designer to describe the metadata, as the protocol designer only needs to write a simple text file, and \protobuf takes cares of creating the necessary code. Another important advantage of \protobuf is its effective way of compressing the data using bit variant techniques that saves storage space and network bandwidth. The protobuf description together with the server and client sides of the protocol are components that the protocol designer needs to provide for any key-value store based on DKVF. These components are shown by dark rectangles in Figure \ref{fig:overall}. We will focus on these components in Section \ref{sec:usage}.

Once a key-value store is ready, we can use it for our applications. An application can be any program (website, mobile application, etc.) that needs to access storage resources through the network. We refer to the entity that provides the storage resources as \textit{storage provider}. The storage provider runs the server side of the key-value store. To do that, the storage provides needs to write a configuration file. This configuration file is an XML file according to \code{Config.xsd} \cite{website}, and describes the cluster and server side parameters. Once the server side is running, different applications can connect to it through the client side of the key-value and use the storage resources. The application developers also need to write a configuration file that specifies servers to connect and client side parameters. The server side and client side configurations are shown by white rectangles in Figure \ref{fig:overall}. These configuration files together with three components that protocol designer needs to write are five components that we need to provide to have a running key-value store based on \name. 

The \code{Application} rectangle in Figure \ref{fig:overall} captures any client program that uses a key-value store server. While the exact application is orthogonal to DKVF, DKVF can be used to provide suitable benchmarks so that application designer can choose the suitable protocol based on these benchmarks. \name relies on YCSB \cite{ycsb} for benchmarking. When we benchmark our key-value store using YCSB, the YCSB client becomes our application in Figure \ref{fig:overall}.

DKVF can be configured to use any storage engine provided the storage developer implements the necessary drivers.
DKVF comes with a driver for Berkeley-BD. We can configure the default storage to be multi-version or single version. In case of multi-version, we have to provide a comparator function to order versions with the same key. DKVF also provides a simple approach for the addition of new storage engine. 
An important question regarding storage is how we want to replicate data. Data replication can be done either by storage engine itself or by the protocol. The default storage delegates data replication to the protocol. This gives the full control of data replication to the protocol designer. However, we can configure DKVF for the case where the storage engine handles data replication.

\section{Creating a Prototype using \name}
\label{sec:usage}
The overall usage of \name is as shown in Figure \ref{fig:overall}. When a designer intends to develop a new protocol, he/she needs to specify three components (shown by dark rectangles in Figure \ref{fig:overall}). These components are 1) metadata description, 2) the server side of the protocol, and 3) the client side of the protocol. We explain each of these components in this section. 

\subsection{Metadata Description}
\label{sec:protobuf}
Describing metadata is done by writing a text file, \code{.proto}, that contains a set of \textit{message} blocks.
You can think of a message as a class or struct in a programming language. Each message has a set of fields. Each field has a type that is either a primitive type like integer, or another message. Any metadata description written for \name must include four messages: 1) \code{Record}, 2) \code{ClientMessage}, 3) \code{ServerMessage}, and 4) \code{ClientReply}. \code{Record} describes records that will be stored in the key-value store. For instance, if we want to store a timestamp with each record, we need to add an \code{int64} field to \code{Record} message to store a 64-bit Java \code{long} variable with each record. \code{ClientMessage} and \code{ServerMessage} describe client and server messages, respectively. \code{ClientReply} describes a response to a client message. 

As an illustration, consider the case where we implement GentleRain protocol \cite{gentleRain} using \name. In GentleRain, each record (i.e., data item) is a tuple as $\langle k, v, ut, sr\rangle$ where $k$ is the key, $v$ is the value, $ut$ is update time (timestamp of the current version), and $sr$ is the ID of the replica where the current version has been written. Listing \ref{list:genRecord} shows the corresponding protobuf description for GentleRain records. Numbers in front of fields are tag numbers that protobuf uses for optimization. Tag numbers must be unique positive integers, and we should assign smaller values to the fields that are used more frequently \cite{protobuf}.


\begin{lstlisting} [label=list:genRecord, caption= {Protobuf description for GentleRain record}, basicstyle=\scriptsize,numberstyle=\ttfamily]
message Record {
    string k = 1; 
    bytes v = 2;
    int64 ut = 3;
    int32 sr = 4;
}
\end{lstlisting}  
In addition to the metadata for the records, we also use protobuf to describe messages that servers/clients send in our protocol. For instance, consider \textsc{GetReq} message in GentleRain \cite{gentleRain}. Each \textsc{GetReq} message has a string to specify the key that we want to read, and an integer for GST value used by the protocol to find a consistent version (see Section \ref{sec:back:consistency}). Listing \ref{list:genGET} shows the necessary protobuf description for \textsc{GetReq} message. Similarly, we can write description for other messages of the protocol.\footnote{The full metadata description needed for GentleRain protocol is provided in the Appendix.}

\begin{lstlisting} [label=list:genGET, caption= Protobuf description for GentleRain \textsc{GetReq} message, 
 basicstyle=\scriptsize,
numberstyle=\ttfamily
]
message GetMessage {
    string k = 1; 
    int64 gst = 2;
}
\end{lstlisting} 
After writing the metadata description, we need to compile it using protobuf. The protobuf will create a Java class that contains all necessary data structures to marshalling/unmarshalling our data. This class is shown by \code{Metadata class} rectangle in Figure \ref{fig:overall}.

\subsection{Server Side Implementation}
\label{sec:server_side}

To implement the server side of a protocol, we need to write a class that extends the abstract class \code{DKVFServer}. \name follows an event-driven approach to define a protocol. Specifically, we can define a protocol  as a set of event handlers. The two main event handlers that will be called by the framework are \code{handleServerMessage} and \code{handleClientMessage} of \code{DKVFServer} class. Inside these two main event handlers, the protocol designer can call detailed event handlers for different events. A protocol can also have other event handlers that do not call by the framework. For instance, GentelRain \cite{gentleRain} and CausalSparatan \cite{causalSpartan} have event handlers that are constantly called at a certain rate. Listing \ref{list:genOverall} shows the overall structure of GentleRainServer class that implements GentleRain on top of DKVF. The body of event handlers is left blank for sake of presentation.

\begin{lstlisting}[label = list:genOverall,
			caption = Overall structure of GetnelRainServer,keywordstyle=\color{blue}\ttfamily,
stringstyle=\color{red}\ttfamily,
     basicstyle=\scriptsize,
   numberstyle=\ttfamily,
          language=java]

public class GentleRainServer extends DKVFServer {
 @Override
 public void handleClientMessage(ClientMessageAgent cma) {
  if (cma.getClientMessage().hasGetMessage()) {
   handleGetMessage(cma);
  } else if (cma.getClientMessage().hasPutMessage()){
   handlePutMessage(cma);
  }
 }

 @Override
 public void handleServerMessage(ServerMessage sm) {
  if (sm.hasReplicateMessage()) {
   handleReplicateMessage(sm);
  } else if (sm.hasHeartbeatMessage()) {
   handleHearbeatMessage(sm);
  } else if (sm.hasVvMessage()) {
   handleVvMessage(sm);
  } else if (sm.hasGstMessage()) {
   handleGstMessage(sm);
  }
 }

 private void handleGetMessage(ClientMessageAgent cma){
  //TODO Handle GET messages here
 }
 private void handlePutMessage(ClientMessageAgent cma){
  //TODO Handle PUT messages here
 }
 private void handleReplicateMessage(ServerMessage sm){
  //TODO Handle Replicate messages here
 }
 void handleHearbeatMessage(ServerMessage sm) {
  //TODO Handle Heartbeat messages here
 }
 void handleVvMessage(ServerMessage sm) {
  //TODO Handle VV messages here
 }
 void handleGstMessage(ServerMessage sm) {
  //TODO Handle GST messages here
 }
}
\end{lstlisting} 


\code{handleServerMessage} receives an object of class \code{ServerMessage} which is created by protobuf from our metadata description explained in Section \ref{sec:protobuf}. 
\code{handleClientMessage} receives an object from class \code{ClientMessageAgent} that includes an object of class \code{ClientMessage} created by protobuf.

While we are processing server or client messages in \code{handleServerMessage} and \code{handleClientMessage}, we may need to send messages to other servers, or send client responses. 
To send a message to another server, the framework provides the convenient \code{sendToServer} method that receives the ID of the destination, and an object of \code{ServerMessage} class. The mapping between server IDs and their actual addresses must be defined in the configuration file. \name takes care of asynchronous reliable FIFO delivery of the message to the destination. Specifically, if the receiver cannot receive the message (e.g., it has crashed, or there is a network partition), \name stores the message and will try to send it later. In the configuration file we can specify the amount of time to wait before resending the message. Also, we can set the capacity of the queue of undelivered messages. If the limit of waiting messages reaches, \name throws an exception. Calling \code{sendToServer} is thread-safe. Thus, protocol designer does not need to worry about concurrency or failure issues. To send the response to the client, the \code{ClientMessageAgent} class provides \code{sendReply} method that allows us to send the response to the client message. 


While we are processing client/server messages, we also need to store or retrieve data from the storage engine. The \code{DKVFServer} class provides methods that can be used for this purpose. Two main methods are \code{read (String k, Predicate<Record> p, List<Record> result)} and \code{insert (String k, Record rec)}. The first method reads all versions of the data item with key $k$ that satisfy predicate $p$. The second method adds record $rec$ to the version chain of the data item with key $k$. The default storage can be configured to be multi-version or single-version. 

Now, as an example, let us consider an implementation of the GET request handler of GentleRain. Algorithm \ref{alg:genGETHandler} shows the pseudocode of this event handler copied from the original paper \cite{gentleRain}.  Listing \ref{list:genHandleGetCode} shows the corresponding necessary code for DKVF. 
First, we update the GST value by calling \code{updateGst} method. This method basically does what Line \ref{alg:line:updateGST} of Algorithm \ref{alg:genGETHandler} does in a thread-safe manner. Next, we call \code{read} of the framework to read the value of the requested data item. We pass a predicate to the read function to find the visible version according to the GentleRain algorithm. Specifically, a version is visible if either it is written locally, or its update time is smaller than GST. Finally, we create \code{ClientReply} message containing the value of the key, and send it to the client by calling \code{sendReply} method of the \code{ClientMessageAgent} object. Note that in Listing \ref{list:genHandleGetCode}, we ignore exception and error handling for sake of presentation.\footnote{The full \name implementation of GentleRain protocol is provided in the Appendix.}

\begin{figure}[t]

\begin{lstlisting}[label = list:genHandleGetCode,
			caption = DKVF implementation of GentleRain GET request handler,			
keywordstyle=\color{blue}\ttfamily,
 stringstyle=\color{red}\ttfamily,
  basicstyle=\scriptsize,
   numberstyle=\ttfamily,
          language=java
         ]
void handleGetMessage(ClientMessageAgent cma) {
 GetMessage gm = cma.getClientMessage().getGetMessage();
 updateGst(gm.getGst()); //Thread-safely update GST
 List Record result = new ArrayList<>();
 StorageStatus ss = read(gm.getKey(), (Record r) -> { 
  if (m == r.getSr() || r.getUt() <= gst.get())
   return true;
  return false;}, result);
 Record rec = result.get(0);
 ClientReply cr = ClientReply.newBuilder().setStatus(true).setGetReply(GetReply.newBuilder().setValue(rec.getValue()).setUt(rec.getUt()).setGst(gst.get())).build();
 cma.sendReply(cr);
}
\end{lstlisting} 
\end{figure}
\begin{algorithm}[t] 
{
\caption{Pseudocode of the GET request handler of GentleRain protocol (copied from \cite{gentleRain})}
\label{alg:genGETHandler}

\begin{algorithmic} [1]

\STATE \textbf{Upon} receive $\langle \textsc{GetReq} \ k, gst\rangle$

\STATE \hspace{3mm} \label{alg:line:updateGST}$GST^m_n \leftarrow max(GST^m_n, gst)$

\STATE \hspace{3mm}  \label{line:get:obtain} obtain latest version $d$ from version chain of key $k$ s.t. $d.sr = m$, or $d.ut < GST^m_n$

\STATE \hspace{3mm} send $\langle\textsc{GetReply} \  d.v, d.ut, GTS^m_n\rangle$ to client
\end{algorithmic}
}
\end{algorithm}


\subsection{Client Side Implementation}

\label{sec:client_side}
To implement the client side of a protocol, we need to extend the client part of the framework. Specifically, we need to write a class that extends class \code{DKVFClient}. When we extend \code{DKVFClient}, we have to implement two abstract methods \code{put} and \code{get} that are the basic PUT and GET operations of a key-value store. These methods are operations that the protocol designer needs to provide for the application developer. The application developer later can use these methods to use the data store (see Figure \ref{fig:overall}). The protocol designer can also add more complex operations for its implementation, but these two methods are required for any implementation.

To process application requests, the client part needs to send client messages and receive responses from the servers. Finding the correct node to send the request is the problem of \textit{service discovery} \cite{book}. \name does not force any service discovery policy, and lets protocol define it. \name, on the other hand, provides convenient ways to send/receive messages to/from servers via their IDs specified in the client configuration file. Specifically, \code{sendToServer(String id, ClientMessage cm)} sends a client message to the server with ID \code{id}, and \code{readFromServer (String id)} reads the response from server with ID \code{id}.

 Now, let us consider client side of PUT operation of GentleRain. Algorithm \ref{alg:client} shows the PUT operation at client side in the GentleRain. Listing \ref{list:genPutCode} shows the corresponding DKVF code. 
To find the correct server to send the PUT request, we call \code{findPartition} function. DKVF \code{Utils} library provides utilities to distribute the keys according to their hash values. The rest of the handler is clear and identical to the pseudocode.

\begin{lstlisting}[label = list:genPutCode,
			caption = DKVF implementation of GentleRain client-side PUT handler,
keywordstyle=\color{blue}\ttfamily,
stringstyle=\color{red}\ttfamily,
basicstyle=\scriptsize,
numberstyle=\ttfamily,
language=java]
public boolean put(String k, byte[] v) {
 ClientMessage cm = ClientMessage.newBuilder().setPutMessage(PutMessage.newBuilder().setDt(dt).setKey(key).setValue(ByteString.copyFrom(value))).build();
 String serverId = findPartition(key) //finds server ID
 sendToServer(serverId, cm);
 ClientReply cr = readFromServer(serverId);
 dt = Math.max(dt, cr.getPutReply().getUt());
 return true;
}
\end{lstlisting}

\begin{algorithm} 

{
\caption{Pseudocode of the PUT handler of the client-side of GentleRain protocol (copied from \cite{gentleRain})}
\label{alg:client}
\begin{algorithmic} [1]
\STATE  \textbf{PUT} (key $k$, value $v$)

\STATE \hspace{3mm} send $\langle \textsc{PutReq} \ k,v,DT_c \rangle$ to server

\STATE \hspace{3mm} receive $\langle \textsc{PutReply} \ ut \rangle$ 

\STATE \hspace{3mm} $DT_c = max (DT_c, ut)$

\end{algorithmic}

}

\end{algorithm}

\section{Benchmarking with YCSB}
\label{sec:ycsb}
YCSB, originally developed by Yahoo!, is a tool for evaluating the performance of key-value or cloud serving stores \cite{ycsb}. To use YCSB, we need to write a YCSB driver that lets YCSB client class use our key-value store. YCSB has a core workload generator.
We can specify different parameters for the core workload generator such as read proportion, insert proportion, value size, number of client threads, number of operations, and so on. Once we specified the workload and driver for YCSB, we can run it to benchmark our system. YCSB gives us different measurements such as throughput and latencies. 

\name comes with a driver for YCSB. Thus, any key-value store written based on \name has its YCSB driver ready. \name also includes a workload generator. The DKVF YCSB workload generator extends YCSB core workload generator by adding new operations such as amplified insert to benchmark the system against query amplification (see Section \ref{sec:back:consistency}).  This feature allows us to evaluate the performance of macro operations that reveal bottlenecks when a query results in multiple operations on the key-value store.

\begin{figure} [h]
\begin{center}
\includegraphics[scale=0.32]{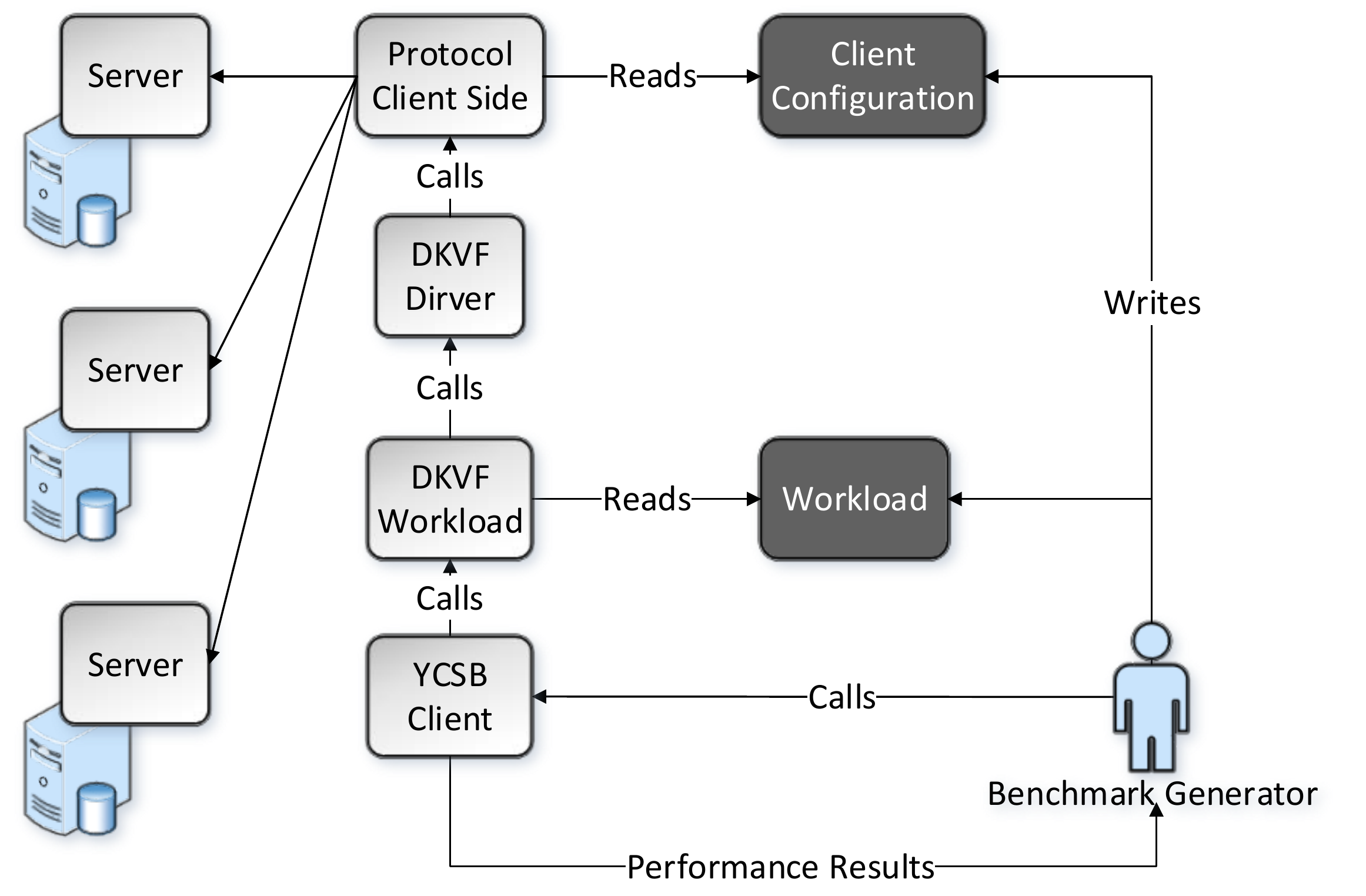}
\caption{Using of YCSB for evaluating a prototype created by DKVF}
\label{fig:ycsb}
\end{center}
\end{figure}

Figure \ref{fig:ycsb} shows the components involving in benchmarking a key-value store created by \name. The person who wants to benchmark the system, referred to as benchmark generator, needs to provide two components shown by dark rectangles in Figure \ref{fig:ycsb}. The first component is the workload properties. The benchmark generator can specify any YCSB core properties for the workload. For benchmarking query amplification, we can specify the amplification factor. The benchmark generator also needs to provide a client configuration file that specifies servers to connect, and other client side parameters (see Section \ref{sec:overview}). The workload generator is also extensible. Specifically, if we want to benchmark an operation that is not included in \name, we need to implement a customized YCSB driver and workload generator. We refer the reader to \cite{ycsb} for details.

\section{Tools}
\label{sec:tools}
In this section, we introduce two tools that help protocol designers to run and benchmark their distributed key-value stores created with \name. These tools can save us a great deal of time and headache in running and benchmarking our systems. 

\subsection{\cm}
\label{sec:cm}
\cm is a command line application to facilitate managing clusters running key-value stores created with \name. It also helps us to run distributed YCSB experiments. Using this tool, we can benchmark our key-value store without directly setting up YCSB; we only need to define our desired workload, and Cluster Manager takes cares of the rest. 

To run a cluster, we need to write a cluster descriptor file. This descriptor file is an XML file according to \code{ClusterDiscriptor.xsd} \cite{website}, and specifies various aspects such as the IP address of the servers, port numbers to listen for incoming client/server messages,  the topology of the servers, and so on. After loading a cluster descriptor file, we can us \cm to start all servers. \cm also enables us to monitor the servers. For instance, we can see if servers have properly started and connected to each other, how much are the network latencies, or how many clients are connected to each server. 

\cm also helps us to test and debug our key-value store. Specifically, after running a cluster we can connect to any server in the cluster and run commands on the servers. For instance, suppose we want to test the convergence of our protocol. We can connect to a replica, and write a value for some key. Next, we can connect to another replica to see if our write has been replicated to the second replica properly. This kind of debugging is very convenient with \cm. \cm uses an instance of the client side of our key-value store to interact with the server. Thus, we need to specify our client class for \cm in the cluster descriptor file. 

After running a cluster, and testing it with \cm, we can conduct an experiment to see how well our protocol performs. We need to write an experiment descriptor file for each experiment. The experiment descriptor file is an XML file  according to \code{ExperimentDescriptor.xsd} \cite{website}, and  specifies experiment related parameters such as how many clients we want to run, what are the addresses of the client machines, each client is connected to which servers, what are the workloads, and so on. We can define a set of experiments for \cm in our descriptor file. After loading an experiment descriptor file, we can use \cm to run the experiment. The \cm conducts the experiments one by one by running YCSB clients, and gather the results from clients. To aggregate the results, \cm provides us with a minimal query language that lets us select measurements we want to aggregate and specify how we want to aggregate them (e.g., taking the average).  

\subsection{Cluster Designer}
\label{sec:cd}
\begin{figure} 
\begin{center}
\includegraphics[scale=0.26]{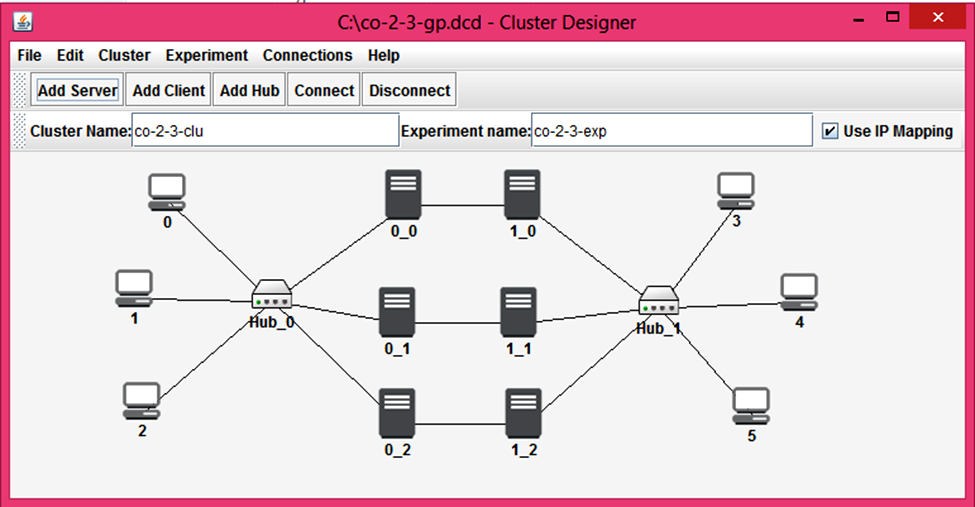}
\caption{The graphical interface of Cluster Designer}
\label{fig:clusterDesigner}
\end{center}
\end{figure}

Although \cm is a convenient tool that can significantly reduce time and headache of  debugging and benchmarking our protocol, writing cluster and experiment descriptor files can be a tedious and of course error-prone task for larger clusters. To solve this issue, we provide Cluster Designer tool. Cluster Designer is a graphical tool that allows us to define our cluster and experiments visually. The tool provides an area where we can add servers and clients. We can connect servers and clients by lines to specify network connections. When we have several components that need to be all connected to each other, we can use hubs to avoid connecting them one-by-one. Figure \ref{fig:clusterDesigner} shows the interface of Cluster Designer. In this network, we have 6 servers and 6 clients. We will talk about this network in more details in \ref{sec:results}. 

We can define default configurations for  servers/clients. We later can tailor default configurations for an individual server/client. After designing our cluster and experiments, we can use Cluster Designer to export descriptor files. We can later use \cm to run our cluster and experiments as explained in Section \ref{sec:cm}.

\section{Experimental Results}

\label{sec:results}

In this section, we present some of the results that we obtained from implementing and evaluating three causal consistency protocols namely COPS \cite{cops} \footnote{We have implemented a simplified version of COPS without garbage collection.}, GentleRain \cite{gentleRain}, and CausalSpartan \cite{causalSpartan} using \name. We also implemented eventual consistency for comparison. Table \ref{tab:lines} shows the number of lines of code that we wrote to implement each of these protocols. For each protocol, we have reported the number of lines that we wrote for server side, client side, and describing our metadata in \code{.proto} file for \protobuf.  Of course, numbers of lines of code is not an accurate indicator, as different people may write the same program in different ways, but we report them here just to give you an estimate of the coding effort that we needed to put to implement these protocols using \name. You can access our implementations in \cite{website}.


\begin{table}

\caption{The number of lines of code that we wrote to implement different protocols with \name.}

\label{tab:lines}

\begin{tabular}{ |l | c | c | c| }

  \hline			

  \textbf{Protocol} & \textbf{Server Side} & \textbf{Client Side} & \textbf{Metadata} \\

  \hline

  Eventual  & 95 & 58 &  32\\

  COPS  & 269 & 84 &  45\\

  GentleRain  & 226 & 61 &  50\\

  CausalSpartan  & 292 & 118 &  53\\

  \hline  

\end{tabular}

\end{table}

Without \name we needed 843 lines to implement CausalSpartan and 769 lines to implement GentleRain. In implementing CausalSpartan and GentleRain without \name, we used Netty \cite{netty} for network communications. The number of lines of code that we needed to implement CausalSpartan and GentleRain with \name is around 40 percent of what we needed to implement them without \name. Note that reducing the number of lines of code is not the only goal of \name. Instead, using \name has all the benefits that we mentioned in the introduction.

We have not implemented COPS and eventual consistency without DKVF. With \name, it took only 2 days to implement COPS based on the description of the protocol in \cite{cops}. Furthermore, all the code developed with DKVF essentially required us to convert the pseudocode in the respective papers into Java and add error handling that is generally omitted in the pseudocode. In this sense, writing the code required with DKVF was straightforward.

\color{black}


\subsection{Experimental Setup}

We consider a replicated and partitioned data store shown in Figure \ref{fig:clusterDesigner}. The data store consists of two replicas. Each replica consists of three partitions. Replica $0$ includes partitions $0\_0$, $0\_1$, and $0\_2$. Replica $1$, on the other hand, consists of partitions $1\_0$, $1\_1$, and $1\_2$. We assume full replication, i.e., each replica has a copy of the entire key space. The key space inside each replica is partitioned among servers. In Figure \ref{fig:clusterDesigner}, we have connected servers inside each replica together with a hub. Partitions are also connected to their peers in the other replica. For servers, we use AWS \code{m3.medium} instances with the following specification:  1 vCPUs, 2.5 GHz, Intel Xeon E5-2670v2, 3.75 GiB memory, 1 x 4 (GB) SSD Storage Capacity.

Connected to each replica, we have a set of clients. We allocate three client machines to run clients. We run 30 threads of YCSB clients on each client machine. All causal consistency protocols that we study here assume locality of traffic, i.e., clients always access one replica. Thus, clients are connected to only one replica as shown Figure \ref{fig:clusterDesigner}. We run clients on \code{c3.large} machines with the following specification: 2 vCPUs, 2.8 GHz, Intel Xeon E5-2680v2, 3.75 GiB memory, 2 x 16 (GB) SSD Storage Capacity. We have used more powerful machines for clients to better utilize our servers.

\subsection{The Effect of Workload on Performance}

The workload of different applications has different characteristics. Some workloads are write-heavy, others like those in data analytics are read-heavy. In this section, we want to study how the characteristics of our workload affect the performance of different consistency protocols. In all experiment, we set the size of the values written by clients to 64 bytes.

\begin{figure} 

\begin{center} 

\includegraphics[scale=0.5]{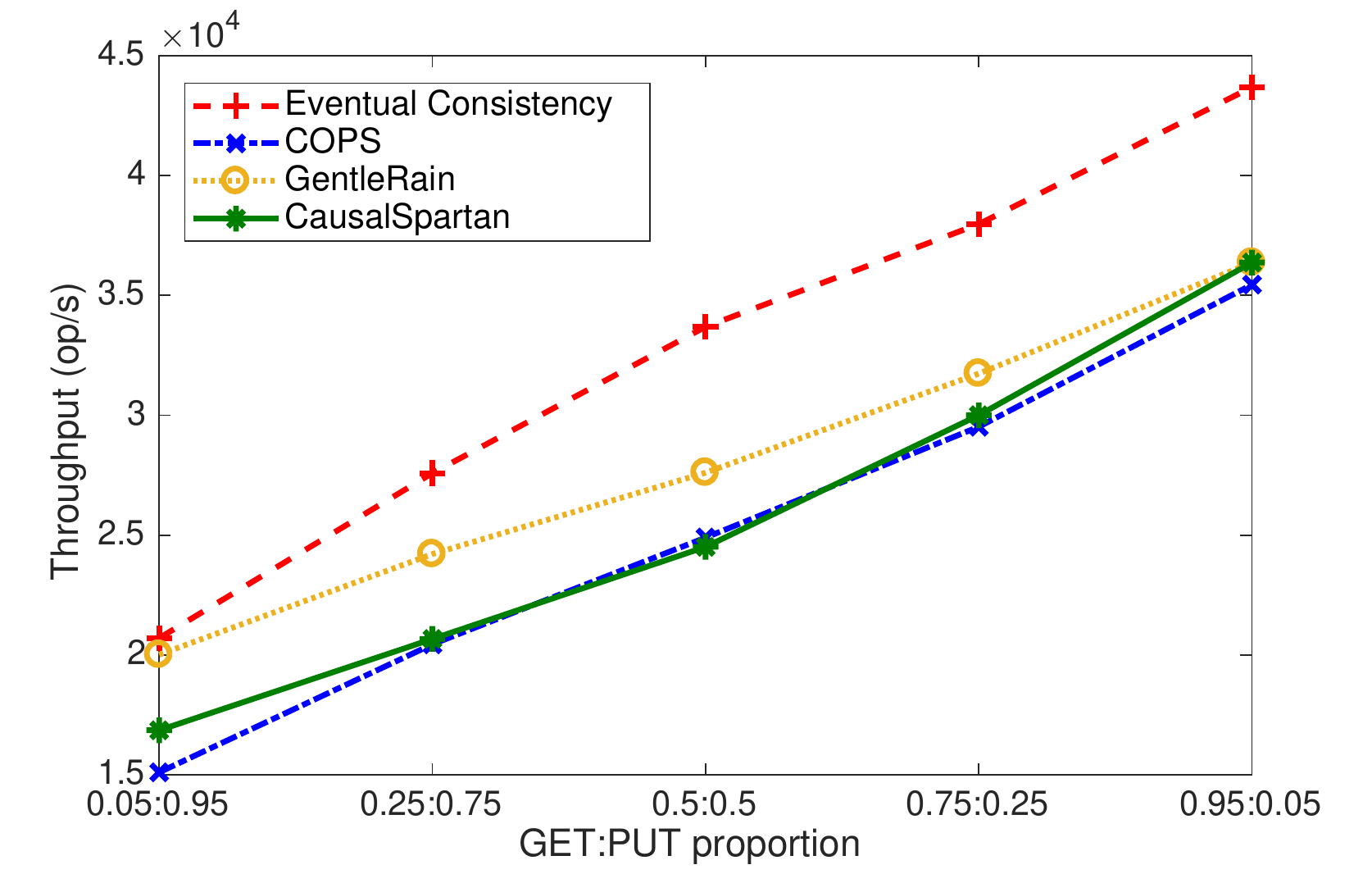}

\caption{Throughput vs. GET:PUT Proportion}

\label{fig:gp_t}

\end{center}

\end{figure}

Figure \ref{fig:gp_t} shows how GET:PUT proportion affects the throughput. As we move from the left side of the plot to its right side, the workload nature changes from write-heavy to read-heavy. The throughputs of all protocol increase as the proportion of GET operations increases. This results confirm previous studies \cite{cops, gentleRain}, and are expected, as GET operations are lighter than PUT.  As expected, eventual consistency has the highest throughput. COPS, on the other hand, has the lowest throughput. This results confirm results published in \cite{gentleRain}, and is due to the overhead of dependency check messages that partitions send to each other to make sure causal dependencies of an update in other partitions are visible (see Section \ref{sec:back:consistency}). 


\begin{figure} 

\begin{center} 

\includegraphics[scale=0.5]{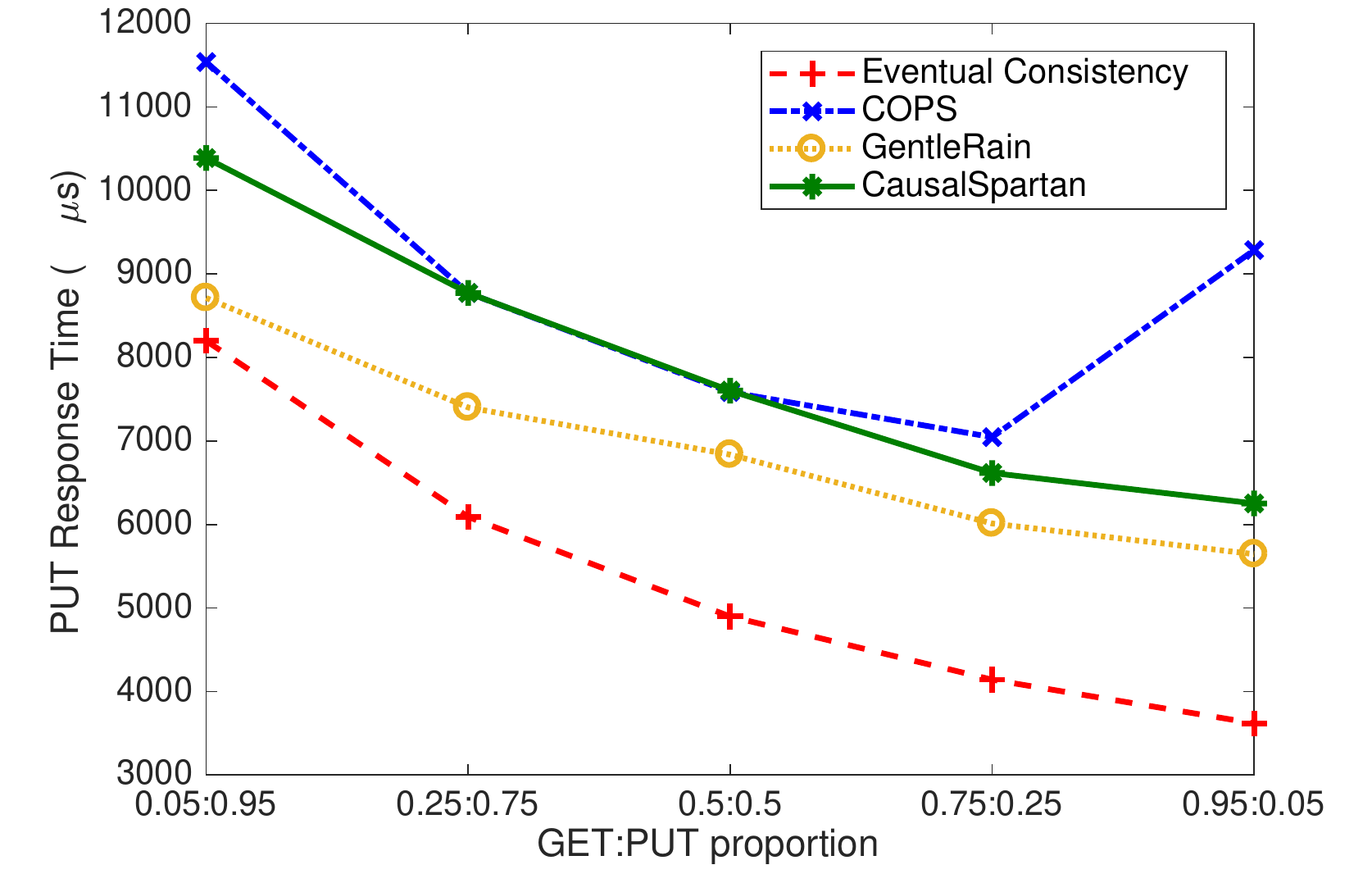}

\caption{Average PUT Response Time vs. GET:PUT Proportion}

\label{fig:gp_put_r}

\end{center}

\end{figure}

Figure \ref{fig:gp_put_r} shows how GET:PUT proportion affects the response time of PUT operations. In all protocols, the response time of PUT operations decreases as we move to read-heavier workloads. This is due to the less load on servers for read-heavier workloads. The eventual consistency has the shortest response time thanks to its minimal metadata. CausalSpartan has more metadata than GentleRain resulting in higher PUT response time. COPS has the highest response time because of its dependency check messages and its explicit dependency tracking approach. Like other protocols, the trend of PUT response time for COPS is decreasing as we move toward read-heavier workloads that can be explained by less load on the machines. However, for 0.05:0.95, the PUT response time increases. This increase can be understood by considering the dependency tracking mechanism of COPS. At point 0.05:0.95, clients read many keys before writing a key. That results in longer dependency lists which make PUT messages heavier to transmit and process. Note that we have implemented a basic version of COPS protocol without client metadata garbage collection. COPS authors suggest a garbage collection mechanism to cope with this problem \cite{cops}.

\begin{figure} 

\begin{center} 

\includegraphics[scale=0.5]{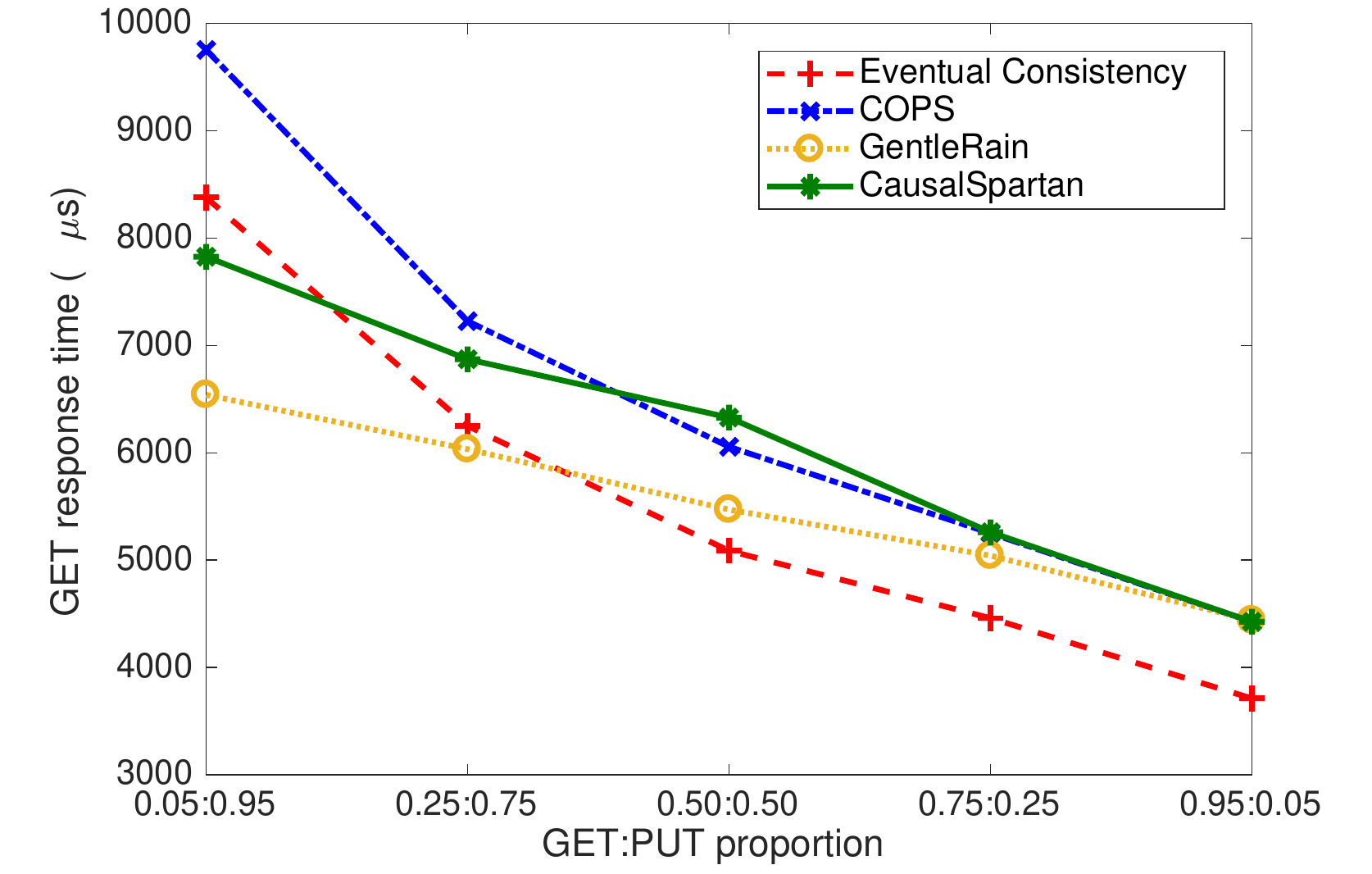}

\caption{Average GET Response Time vs. GET:PUT Proportion}

\label{fig:gp_get_r}

\end{center}

\end{figure}

Figure \ref{fig:gp_get_r} shows how GET:PUT proportion affects the response time of GET operations. Like the case of PUT operations, the response time of GET operations also decreases, as we move towards read-heavier workloads. 
It is interesting that GentleRain and CausalSpartan have a lower response time for GET operations comparing to the eventual consistency for write-heavy workloads. This can be explained by the synchronization that occurs between threads in GentleRain and CausalSpartan. Specifically, there is a contention between threads while performing PUT operations in GentleRain/CausalSpartan. This contention occurs for obtaining a lock that we used to guarantee updates with smaller timestamps are replicated to other nodes before updates with higher timestamps. This increases the PUT response time that results in lower overall throughput of GentleRain/CausalSpartan for write-heavy workloads. While threads serving PUT operations are waiting for synchronization, the server can handle GET operations. On the other hand, in the eventual consistency, there is no competition between PUT operations. Thus, there are more active threads serving PUT operations leading to higher competition over CPU that finally results in higher GET response time comparing to GentleRain/CausalSpartan. Note that this happens for write-heavy workloads with low GET proportion. Therefore, the eventual consistency still has the highest overall throughout in all cases (See Figure \ref{fig:gp_t}).

\subsection{The Effect of Query Amplification}

In this section, we study the effect of query amplification on the performance of the system. In this section, we only consider one replica consisting of three partitions. We consider a workload that purely consists of amplified insert operations. Each amplified insert consists of several internal PUT operations. The number of internal PUT operations is defined by the amplification factor.

\begin{figure} 

\begin{center} 

\includegraphics[scale=0.5]{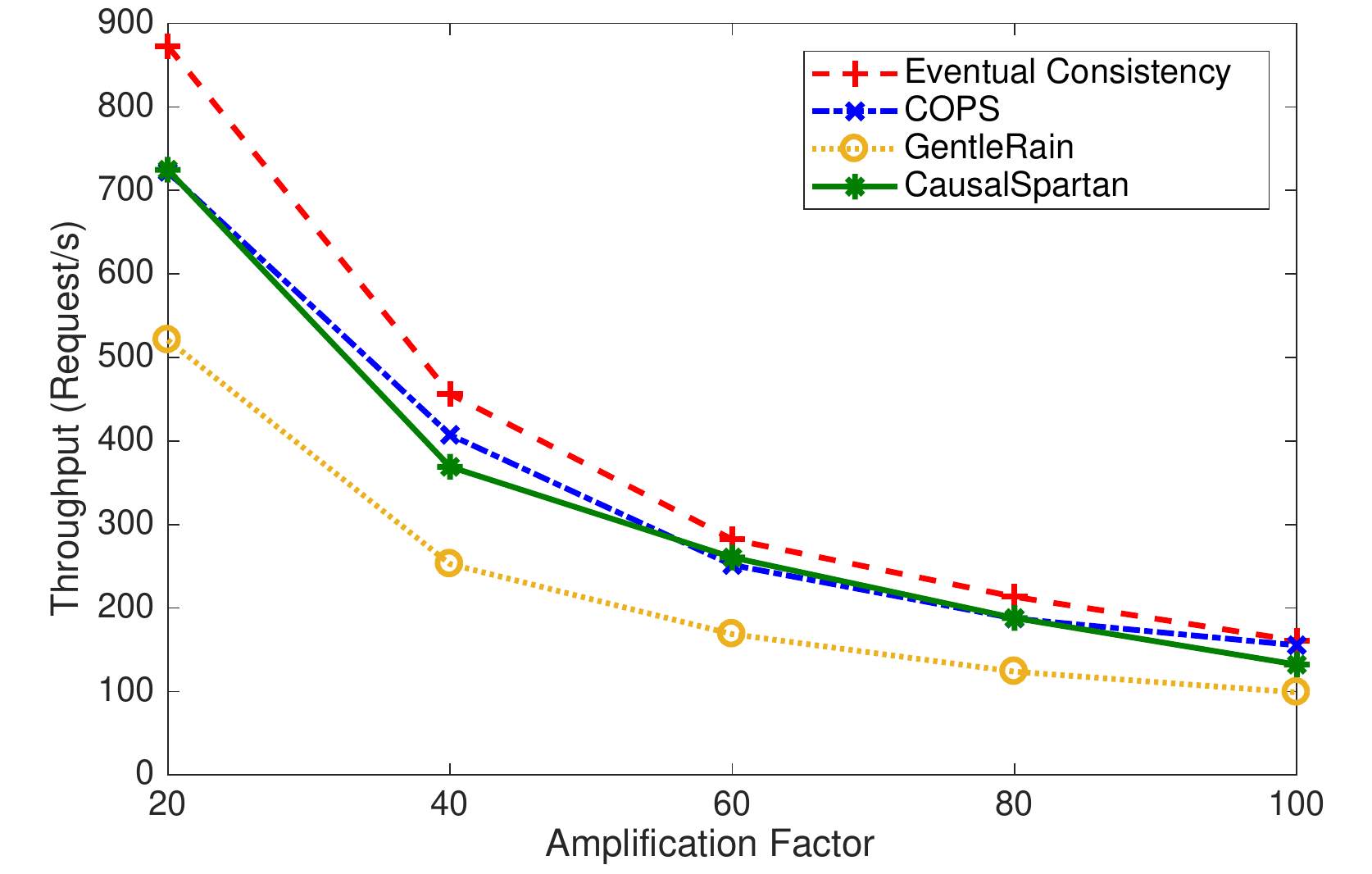}

\caption{Throughput vs. Amplification Factor}

\label{fig:amp_t}

\end{center}

\end{figure}

\begin{figure} 

\begin{center} 

\includegraphics[scale=0.5]{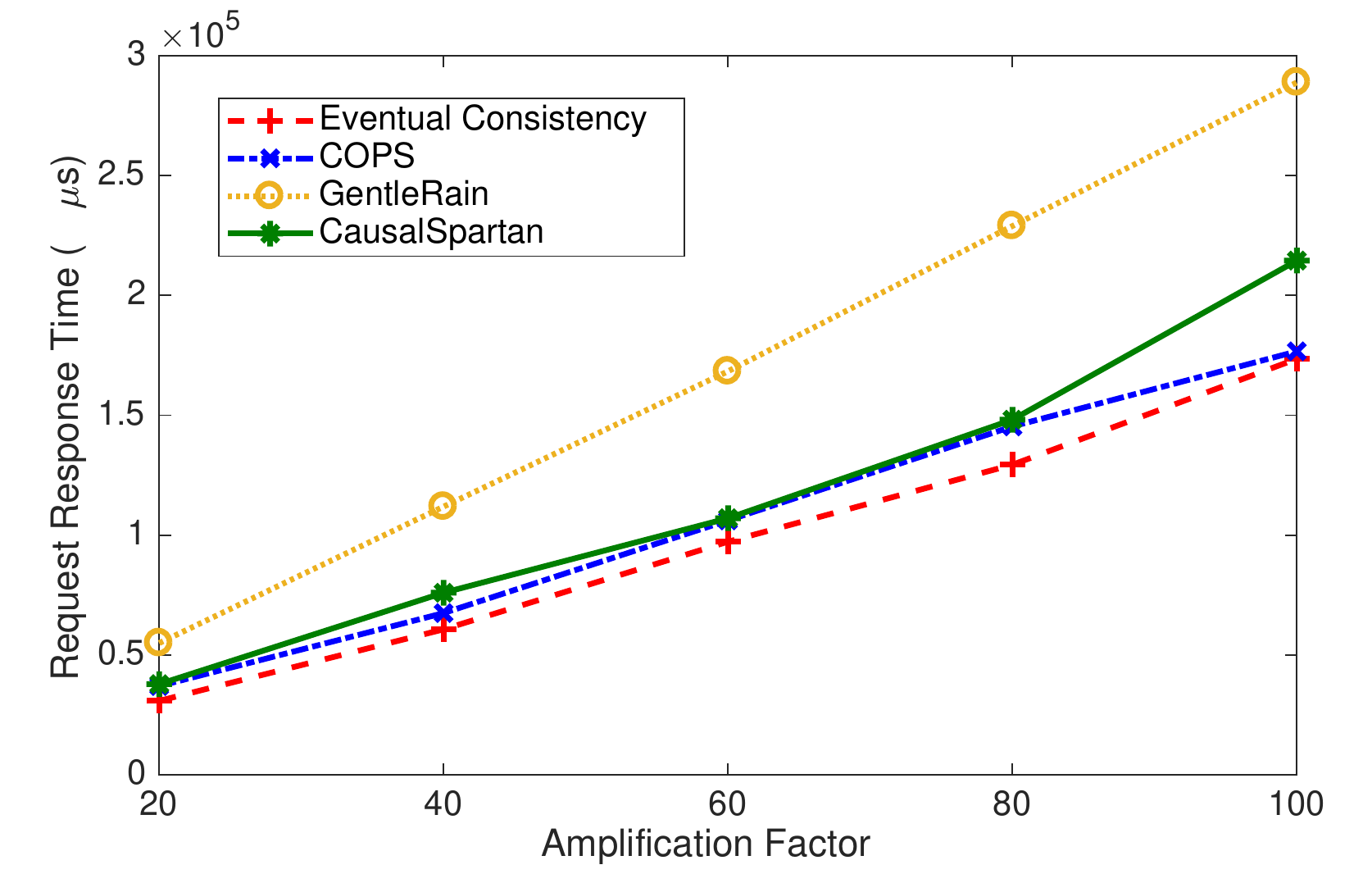}

\caption{Request Response Time vs. Amplification Factor}

\label{fig:amp_r}

\end{center}

\end{figure}

Figure \ref{fig:amp_t} shows the effect of amplification factor on the client request throughput. Note that this throughput represents the number of client macro operations (not individual PUT operations) that are served in one second. As the amplification factor increases, the throughput of all protocol decreases which is expected, as requests with higher amplification factor include more internal operations which mean more job to do for each request. The eventual consistency has the highest throughput. The pure-write workload is an ideal write scenario for COPS, as dependency lists have at most one entry. Thus, the throughput of COPS is the highest after eventual consistency for this scenario. GentleRain has the lowest throughput. That is due to the delay that GentleRain imposes on PUT operations in case of clock skew between servers. Note that we synchronized the physical clocks of the system with NTP \cite{ntp}, but the effect of clock skew still shows up in the results. These results confirm previous results presented in \cite{causalSpartan}. CausalSpartan has higher throughput than GentleRain, as CausalSpartan eliminates the need for the delay before PUT operations by utilizing HLCs instead of physical clocks\cite{causalSpartan}. 
Figure \ref{fig:amp_r} shows the request response time for different protocols. Again, because of delays that GentelRain forces on PUT operations, request response time has the highest value for GentleRain.

\section{Conclusion and Future Work}
\label{sec:con}

 In this paper, we introduced \name which is a framework for rapid prototyping and benchmarking distributed key-value stores. It streamlines the evaluation of the performance of consistency protocols for distributed key-value store. To show the effectiveness of our framework, we implemented four consistency protocols using \name. Thanks to the convenience of \name, we were able to implement each of these protocols in less than 2 days. We were able to implement CausalSpartan and GentleRain with significantly less effort than our previous implementations without \name. 
Note that in implementing CausalSpartan and GentleRain without \name, we used Netty \cite{netty} that helped us for network communications. Although frameworks like Netty streamline network programming, to implement a distributed key-value store still we have to write code for many parts that are independent of the logic of the protocol. \name and its toolset provide a more straightforward framework that is specialized to develop distributed key-value stores.
We believe other groups can also benefit from \name by reducing the necessary implementation efforts.

\name relies on YCSB for benchmarking. The toolset that comes with the framework helps protocol designers to easily evaluate their prototype. 
We evaluated the prototypes that we developed by \name using the tools provided by the framework.  Our results are consistent with what has been previously reported in the literature, and also with our previous results from prototypes that we developed without \name.

We can use any storage systems as the storage engine for the key-value stores that we develop with \name. This enables protocol designers to flexibility change their storage engine without touching the implementation of  their consistency protocol. To use a given storage system with \name, we need to write a driver for it that enables \name to interact with it. \name comes with a driver for Berkeley-DB. Writing drivers for other storage systems is part of the future work. Also, \name is designed to be extensible so that these drivers can be easily added by others.

\bibliography{bib}
\newpage

 \appendix
\begin{lstlisting} [label=list:protobuf, caption= DKVF protobuf description for GentleRain protocol
]
message Record {
	string key = 1; 
	bytes value = 2; 
	int64 ut = 3;
}

message GetMessage {
	string key =1;
}

message PutMessage {
	string key =1;
	bytes value = 2;
}

message ClientMessage {
	oneof message_type {
		GetMessage get_message = 1;
		PutMessage put_message = 2;
	}
}

message GetReply{
	bytes value = 1;
}

message PutReply{
	int64 ut = 1;
}

message ClientReply {
	oneof message_type{
		GetReply get_reply= 1;
		PutReply put_reply= 2;
	}
}

message ReplicateMessage {
	Record rec = 1;
}

message ServerMessage {
	ReplicateMessage replicate_message = 1;
}
\end{lstlisting}

\begin{lstlisting}[label = list:server_part,
			caption = Server side for GentleRain protocol,
            float=*,
           keywordstyle=\color{blue}\ttfamily,
           stringstyle=\color{red}\ttfamily,
           language=java]
public class GentleRainServer extends DKVFServer {
	AtomicLong gst = new AtomicLong(0);
	int dcId;// datacenter id
	int pId; // partition id
	int numOfDatacenters;
	int numOfPartitions;
	// GST computation
	ArrayList<AtomicLong> vv;
	HashMap<Integer, List<Long>> childrenVvs;
	// Tree structure
	int parentPId;
	List<Integer> childrenPIds;
	// intervals
	int heartbeatInterval;
	int gstComutationInterval;
	// Heartbeat
	long timeOfLastRepOrHeartbeat;

	Object putLock = new Object(); 

	public GentleRainServer(ConfigReader cnfReader) {
		super(cnfReader);
		HashMap<String, List<String>> protocolProperties = cnfReader.getProtocolProperties();

		dcId = new Integer(protocolProperties.get("dc_id").get(0));
		pId = new Integer(protocolProperties.get("p_id").get(0));

		parentPId = new Integer(protocolProperties.get("parent_p_id").get(0));
		childrenPIds = new ArrayList<Integer>();
		if (protocolProperties.get("children_p_ids") != null) {
			for (String id : protocolProperties.get("children_p_ids")) {
				childrenPIds.add(new Integer(id));
			}
		}

		numOfDatacenters = new Integer(protocolProperties.get("num_of_datacenters").get(0));
		numOfPartitions = new Integer(protocolProperties.get("num_of_partitions").get(0));

		heartbeatInterval = new Integer(protocolProperties.get("heartbeat_interval").get(0));
		gstComutationInterval = new Integer(protocolProperties.get("gst_comutation_interval").get(0));

		vv = new ArrayList<>();
		ArrayList<Long> allZero = new ArrayList<>();
		for (int i = 0; i < numOfDatacenters; i++) {
			vv.add(i, new AtomicLong(0));
			allZero.add(new Long(0));
		}
		childrenVvs = new HashMap<>();
		for (int cpId: childrenPIds){
			childrenVvs.put(cpId, allZero);
		}
		// Scheduling periodic operations
		ScheduledExecutorService heartbeatTimer = Executors.newScheduledThreadPool(1);
		ScheduledExecutorService gstComputationTimer = Executors.newScheduledThreadPool(1);

		heartbeatTimer.scheduleAtFixedRate(new HeartbeatSender(this), 0, heartbeatInterval, TimeUnit.MILLISECONDS);

\end{lstlisting}

\begin{lstlisting}[
            float=*,
           keywordstyle=\color{blue}\ttfamily,
           stringstyle=\color{red}\ttfamily,
           language=java,
           firstnumber=57]
		gstComputationTimer.scheduleAtFixedRate(new GstComputation(this), 0, gstComutationInterval, TimeUnit.MILLISECONDS);
	}
	@Override
	public void handleClientMessage(ClientMessageAgent cma) {
		if (cma.getClientMessage().hasGetMessage()) {
			handleGetMessage(cma);
		} else if (cma.getClientMessage().hasPutMessage()) {
			handlePutMessage(cma);
		}
	}
	@Override
	public void handleServerMessage(ServerMessage sm) {
		if (sm.hasReplicateMessage()) {
			handleReplicateMessage(sm);
		} else if (sm.hasHeartbeatMessage()) {
			handleHearbeatMessage(sm);
		} else if (sm.hasVvMessage()) {
			handleVvMessage(sm);
		} else if (sm.hasGstMessage()) {
			handleGstMessage(sm);
		}
	}
	private void handleGetMessage(ClientMessageAgent cma) {
		GetMessage gm = cma.getClientMessage().getGetMessage();
		updateGst(gm.getGst());
		List<Record> result = new ArrayList<>();
		StorageStatus ss = read(gm.getKey(), isVisible, result);
		ClientReply cr = null;
		if (ss == StorageStatus.SUCCESS) {
			Record rec = result.get(0);
			cr = ClientReply.newBuilder().setStatus(true).setGetReply(GetReply.newBuilder().setValue(rec.getValue()).setUt(rec.getUt()).setGst(gst.get())).build();
		} else {
			cr = ClientReply.newBuilder().setStatus(false).build();
		}
		cma.sendReply(cr);
	}
	Predicate<Record> isVisible = (Record r) -> {
		protocolLOGGER.finer(MessageFormat.format("record ut= {0}, Current GST={1}", r.getUt(), gst.get()));
		if (dcId == r.getSr() || r.getUt() <= gst.get())
			return true;
		return false;
	};
	private void updateGst(long sample) {
		while (true) {
			long curMax = gst.get();
			if (curMax >= sample) {
				break;
			}
			boolean setSuccessful = gst.compareAndSet(curMax, sample);
			if (setSuccessful) {
				break;
			}
		}
	}
	private void handlePutMessage(ClientMessageAgent cma) {
		PutMessage pm = cma.getClientMessage().getPutMessage();
		long sleepTime = pm.getDt() - System.currentTimeMillis();
		try {
\end{lstlisting}
\begin{lstlisting}[
            float=*,
           keywordstyle=\color{blue}\ttfamily,
           stringstyle=\color{red}\ttfamily,
           language=java,
           firstnumber=115]
          	 if (sleepTime > 0){
				Thread.sleep(sleepTime);
				protocolLOGGER.info("Sleeping for " + sleepTime);
			}
		} catch (InterruptedException e) {
			protocolLOGGER.severe("Failed to delay write operation.");
		}
		vv.get(dcId).set(System.currentTimeMillis());
		Record rec = null; 
		
		synchronized (putLock) {
			rec = Record.newBuilder().setValue(pm.getValue()).setUt(vv.get(dcId).get()).setSr(dcId).build();
			sendReplicateMessages(pm.getKey(),rec); 
		}
		StorageStatus ss = insert(pm.getKey(), rec);
		ClientReply cr = null;
		if (ss == StorageStatus.SUCCESS) {
			cr = ClientReply.newBuilder().setStatus(true).setPutReply(PutReply.newBuilder().setUt(rec.getUt())).build();
		} else {
			cr = ClientReply.newBuilder().setStatus(false).build();
		}
		cma.sendReply(cr);
	}
	private void sendReplicateMessages(String key, Record recordToReplicate) {
		ServerMessage sm = ServerMessage.newBuilder().setReplicateMessage(ReplicateMessage.newBuilder().setDcId(dcId).setKey(key).setRec(recordToReplicate)).build();
		for (int i = 0; i < numOfDatacenters; i++) {
			if (i == dcId)
				continue;
			String id = i + "_" + pId;

			protocolLOGGER.finer(MessageFormat.format("Sendng replicate message to {0}: {1}", id, sm.toString()));
			sendToServerViaChannel(id, sm);
		}
	}
	private void handleReplicateMessage(ServerMessage sm) {
		protocolLOGGER.finer(MessageFormat.format("Received replicate message: {0}", sm.toString()));
		int senderDcId = sm.getReplicateMessage().getDcId();
		Record d = sm.getReplicateMessage().getRec();
		insert(sm.getReplicateMessage().getKey(), d);
		vv.get(senderDcId).set(d.getUt());
	}
	void handleHearbeatMessage(ServerMessage sm) {
		int senderDcId = sm.getHeartbeatMessage().getDcId();
		vv.get(senderDcId).set(sm.getHeartbeatMessage().getTime());
	}
	void handleVvMessage(ServerMessage sm) {
		int senderPId = sm.getVvMessage().getPId();
		List<Long> receivedVv = sm.getVvMessage().getVvItemList();
		childrenVvs.put(senderPId, receivedVv);
	}
	void handleGstMessage(ServerMessage sm) {
		Long receivedGst = sm.getGstMessage().getGst();
		gst.set(receivedGst);
		sm = ServerMessage.newBuilder().setGstMessage(GSTMessage.newBuilder().setGst(gst.get())).build();
		sendToAllChildren(sm);
	}
	void sendToAllChildren(ServerMessage sm) {
		for (Map.Entry<Integer, List<Long>> child : childrenVvs.entrySet()) {
			int childId = child.getKey();
			sendToServerViaChannel(dcId + "_" + childId, sm);
		}
	}
}
\end{lstlisting}
\begin{lstlisting}[
            float=*,
           keywordstyle=\color{blue}\ttfamily,
           stringstyle=\color{red}\ttfamily,
           language=java,
           firstnumber=178]
public class GstComputation implements Runnable {
	GentleRainServer server;
	public GstComputation(GentleRainServer server) {
		this.server = server;
	}

	@Override
	public void run() {
		//take minimum of all childrens 
		List<Long> minVv = new ArrayList<Long>();
		for (AtomicLong v : server.vv) {
			minVv.add(v.get());
		}
		for (Map.Entry<Integer, List<Long>> childVv : server.childrenVvs.entrySet()) {
			for (int i = 0; i < childVv.getValue().size(); i++) {
				if (minVv.get(i) > childVv.getValue().get(i))
					minVv.set(i, childVv.getValue().get(i));
			}
		}

		//if the node is parent it send Gstmessage to its children
		ServerMessage sm = null;
		if (server.parentPId == server.pId) {
			Long newGst = Long.MAX_VALUE;
			for (Long l : minVv) {
				newGst = Math.min(l, newGst);
			}
			server.gst.set(newGst);
			sm = ServerMessage.newBuilder().setGstMessage(GSTMessage.newBuilder().setGst(server.gst.get())).build();
			server.sendToAllChildren(sm);
		}
		//if the node is not root, it send vvMessage to its parent.
		else {
			VVMessage vvM = VVMessage.newBuilder().setPId(server.pId).addAllVvItem(minVv).build();
			sm = ServerMessage.newBuilder().setVvMessage(vvM).build();
			server.sendToServerViaChannel(server.dcId + "_" + server.parentPId, sm);
		}
	}
}

public class HeartbeatSender implements Runnable {
	GentleRainServer server;
	public HeartbeatSender(GentleRainServer server) {
		this.server = server;
	}
	@Override
	public void run() {
		long ct = System.currentTimeMillis(); 
		if (ct > server.timeOfLastRepOrHeartbeat + server.heartbeatInterval){
			server.vv.get(server.dcId).set(ct);
			ServerMessage sm = ServerMessage.newBuilder().setHeartbeatMessage(HeartbeatMessage.newBuilder().setDcId(server.dcId).setTime(ct)).build();
			for (int i = 0; i < server.numOfDatacenters; i++) {
				if (i == server.dcId)
					continue;
				server.sendToServerViaChannel(i + "_" + server.pId, sm);
			}
		}
	}
}

\end{lstlisting} 

\end{document}